\documentclass[11pt,a4paper]{article}

\usepackage[margin=2.3cm]{geometry}
\usepackage{amsmath,amssymb,amsfonts}
\usepackage{graphicx}
\usepackage{subcaption}
\usepackage{float}
\usepackage{lineno}
\usepackage{hyperref}

\linenumbers

\usepackage{amsmath}
\usepackage{amssymb}
\usepackage{amsfonts}

\usepackage{lineno}
\linenumbers 

\usepackage{subcaption}
\usepackage{float}
\usepackage{setspace}

\newcommand{\bb}{\begin{equation}}
\newcommand{\ee}{\end{equation}}
\newcommand{\bega}{\begin{eqnarray}}
\newcommand{\ega}{\end{eqnarray}}
\newcommand{\begae}{\begin{eqnarray*}}
\newcommand{\egae}{\end{eqnarray*}}

\newcommand{\dis}{\displaystyle}

\newcommand{\om}{\omega}

\title{Structured light sheets}

\author{
Michel Zamboni-Rached$^{1,*}$,
Nima Asoudegi$^{2}$,
Joel Alcídio Varela Mendonça$^{1}$,
and Mo Mojahedi$^{2}$\\[0.3cm]
{\small $^{1}$University of Campinas (UNICAMP), Campinas, SP, Brazil}\\
{\small $^{2}$University of Toronto, Toronto, ON, Canada}\\
{\small $^{*}$mzamboni@decom.fee.unicamp.br}
}

\date{}

\begin{document}

\nolinenumbers

\onehalfspacing

\maketitle


\begin{abstract}

In this work, we present a simple, exact, and fully analytical method for generating light sheets parallel to the propagation direction, with amplitude and phase envelopes structured on demand. We validate the approach theoretically and experimentally by imprinting images onto light sheets, and we compare the theoretical performance with that obtained using an alternative strategy based on arrays of Frozen Waves (FWs). In this context, the proposed method provides a more direct and flexible control of the field envelopes on the light sheets, resulting in higher-fidelity reconstructions than those achieved with FW-based approaches. The method thus offers a versatile framework for structured light-sheet generation, with potential applications in optical manipulation, microscopy, and 3D holographic imaging.
\end{abstract}

\section{Introduction}

While structured optical beams \cite{fw1,fw2,fw3,fw4,exp1,livro1,livro2,Mateus,Yousuf1,Yousuf2,AhmedForbes,AhmedMichelMo,Carving1,Forbes1,Forbes2,Forbes3,Zhang,Carving2,AhmedCapasso1,Chen,AhmedCapasso2,Park,AhmedCapasso3,Shaltout,Ren}
 have been extensively studied over the last two decades, leading to significant theoretical and applied advances, it is only more recently that the concept of structured optical fields has been considered in the context of light sheets \cite{Vett,Bi,Leo1,Leo2,Nature}.

In this work, we revisit a recent method \cite{PW} based on a suitable superposition of plane waves (PWs), here referred to as the PW method, originally developed for the construction of structured optical beams along curved trajectories. By introducing an essential adjustment in the approach, which enables the extraction of the full spectral content compatible with unidirectional propagation, we optimize the method for the generation of highly complex structured light sheets parallel to the propagation direction, with amplitude and phase envelopes tailored on demand.

To demonstrate the effectiveness of this approach, we use the method to generate images on a single light sheet, as well as on three parallel light sheets, and compare the results with those obtained using an alternative approach based on arrays of Frozen Wave (FW) beams \cite{fw1,fw2,fw3,Leo1,Leo2,Nature}. Although the generation of a two-dimensional image, when regarded solely as an intensity envelope on the light-sheet plane, does not require explicit phase control, this comparison provides a clear and objective benchmark, revealing the superior fidelity achieved by the PW method. Finally, we experimentally validate the approach by generating the theoretically predicted stamped light sheets using computer-generated holograms implemented on a Spatial Light Modulator (SLM).

From a more fundamental perspective, an important distinction between the two approaches emerges when considering a single planar light sheet. In this case, the PW method yields a set of orthogonal spatial modes on the sheet (although truncated to exclude evanescent components), allowing independent and faithful synthesis of complex amplitude and phase distributions. In contrast, superpositions of Frozen Waves, as employed in previous works on light sheets, do not form an orthogonal basis on the plane, which inevitably results in undesired interference among the composing fields. This structural difference explains the higher fidelity achieved by the PW method, even in situations where only amplitude information is encoded, such as image generation.

Due to its high efficiency, low computational overhead, mathematical simplicity, exact and analytical nature, as well as its straightforward generalization to the vector case, we believe that the method proposed in this work constitutes a significant contribution to the field of light sheets. Furthermore, compared with FW-based constructions and iterative optimization techniques such as Gerchberg-Saxton and Wirtinger holography, the proposed framework offers a highly scalable route toward the synthesis of complex three-dimensional structured optical fields through arrays of parallel light sheets, enhancing its applicability in areas such as optical manipulation, microscopy, 3D holographic imaging, and beyond.


\section{The method}

The aim of this paper is to construct an optical field predominantly propagating along the positive $z$ direction, without counterpropagating components, concentrated over planes $y=y_p$ ($p=1,2,3,..,P$) and whose envelopes $\psi^{(p)}(x,y,z)$ can assume, on those planes and within $-L_x/2 \leq x \leq L_x/2$, $0 \leq z \leq L_z$, values of amplitude and phase given by complex morphological functions $F^{(p)}(x,z)$ chosen at will. In summary, the mentioned optical field is given by a superposition of parallel light sheets, each concentrated on its corresponding plane ($y=y_p$).

This can be achieved by considering a superposition of plane waves endowed with wave vectors $\mathbf{k}_{mn} = k_{xm}\hat{x}+k_{zn}\hat{z}+k_{ymn}\hat{y}$, where $k_{xm}=2\pi m /L_x$, $k_{zn}= Q + 2\pi n /L_z$ and $k_{ymn} = \pm \sqrt{k^2 - k_{xm}^2 - k_{zn}^2}$, with $0<Q<k=n_{RI}\omega/c$, $n_{RI}$ denoting the refractive index of the medium, $\omega$ the angular frequency, $c$ the speed of light, and $Q$ a constant to ensure the predominant propagation along the positive $z$ direction.

Mathematically, we write the (scalar) solution of the optical field as

\bb \Psi(x,y,z,t) = e^{-i\om t} e^{i Q z} \, \dis{\sum_{p=1}^{P}\psi^{(p)}(x,y,z)} \,\,, \label{Psi}  \ee
where each envelope $\psi^{(p)}(x,y,z)$ is given by

\bb
\begin{array}{l}
\psi^{(p)}(x,y,z)
= \dis{\frac{1}{2}} \, \dis{\sum_{n=-N}^{N} \, \sum_{m=-M_n}^{M_n}}
A_{mn}^{(p)} \, e^{i\frac{2\pi}{L_z}nz}\,e^{i\frac{2\pi}{L_x}mx}
\left(
e^{+i\sqrt{k^2 - (2\pi m/L_x)^2 - (Q + (2\pi n/L_z))^2}\,(y-y_p)}\right.

\\
\\

 + \left. e^{-i\sqrt{k^2 - (2\pi m/L_x)^2 - (Q + (2\pi n/L_z))^2}\,(y-y_p)}\right)

\\
\\

= \dis{\sum_{n=-N}^{N} \, \sum_{m=-M_n}^{M_n}}
A_{mn}^{(p)} \, e^{i\frac{2\pi}{L_z}nz}\,e^{i\frac{2\pi}{L_x}mx}
\cos\!\left(
\sqrt{k^2 - (2\pi m/L_x)^2 - (Q + (2\pi n/L_z))^2}\,(y-y_p)
\right)
\,,
\end{array}
\label{psi}
\ee where $A_{mn}^{(p)}$ are the complex amplitudes, which are still unknown, and $N=\text{floor}(L_z(k-Q)/2\pi)$, $M_n=\text{floor}(L_x\sqrt{k^2-(Q+2\pi n/L_z)^2}\,/2\pi)$.

The value of $N$ in Eq.~(\ref{psi}) is obtained from the condition
$0 \leq k_{zn} = Q + (2\pi n)/L_z \leq k$,
taking into account that we generally consider $Q \lesssim k$.
The value of $M_n$ is determined based on the fact that, for a plane wave
with a given $k_z$, the maximum allowed transverse component is
$(k_x)_{\max} = \sqrt{k^2 - k_z^2}$.
In Eq.~(\ref{psi}), the limits of the summation over $k_{xm}$,
which depend on the value of $n$ used for $k_{zn}$,
constitute an improvement over the method developed in \cite{PW},
enabling higher resolution of the structured light sheets,
as will be shown in the next section.

As stated at the beginning of this section, one of our goals is to obtain
$\psi^{(p)}(x,y=y_p,z) \approx F^{(p)}(x,z)$
over the plane $y=y_p$ and within
$-L_x/2 \leq x \leq L_x/2$ and $0 \leq z \leq L_z$,
where $F^{(p)}(x,z)$ is a complex function, here referred to as the
morphological function, chosen on demand.
From Eq.~(\ref{psi}), this objective can be achieved by assigning the
complex amplitudes $A_{mn}^{(p)}$ as the Fourier coefficients of
$F^{(p)}(x,z)$, namely,

\bb A_{mn}^{(p)} = \frac{1}{L_x}\,\frac{1}{L_z} \, \dis{\int_{-L_x/2}^{L_x/2}\,\int_{0}^{L_z}} \, F^{(p)}(x,z)\,e^{-i\frac{2\pi}{L_x}mx}\,e^{-i\frac{2\pi}{L_z}nz}\,{\rm d}z\,{\rm d}x \label{Amn}   \ee

It is noteworthy that each resulting structured envelope field,
as described by Eq.~(\ref{psi}),
is concentrated around its corresponding plane $y=y_p$,
with a thickness that depends on the morphological function
$F^{(p)}(x,z)$, particularly in cases where its phase varies spatially.
In such situations, the transverse width $\Delta y_p$ of the $p$-th
light sheet may become a function of the coordinates $x$ and $z$
on the plane $y=y_p$.
In cases where the phase of the morphological function remains constant
or varies weakly (such as in the case of an image),
the thickness can be approximately estimated as
$\Delta y_p = \Delta y \approx \pi/(2\sqrt{k^2-Q^2})$.
This result shows that the parameter $Q$ not only governs the paraxiality
of the resulting envelope fields, but also determines the thickness of
each light sheet.

At this point, however, an important observation should be made.
Although the approximate estimate for the structured light sheet thickness may suggest that it
can be made arbitrarily thin by an appropriate choice of the parameter $Q$,
it should be emphasized that this quantity does not correspond to a
monotonic transverse decay length of the field intensity.
In fact, whenever a structured light sheet is designed to resist diffraction
over extended propagation distances, an oscillatory decay of the intensity
along the direction perpendicular to the light-sheet plane is expected to occur.
In this sense, the estimate for $\Delta y$ should be interpreted as an
approximate measure of the distance from the light-sheet plane to the first
transverse near-zero of the field,
in close analogy with the usual definition of the spot radius of a
zero-order Bessel beam.

As a consequence of this oscillatory decay, the field intensity does not
vanish monotonically away from the plane $y=y_p$, but may exhibit residual oscillatory contributions at transverse distances
larger than $\Delta y$. When multiple parallel light sheets are considered, this behavior may lead to
a certain degree of crosstalk between adjacent sheets.
It is important to stress, however, that this effect is not a limitation
of the present method, but rather a general feature associated with
diffraction-resistant structured optical fields.

Another point that deserves attention concerns the discrete nature of the
plane-wave superposition employed in Eq.~(\ref{psi}).
The discretization of the transverse and longitudinal wave-vector components
$k_{xm}$ and $k_{zn}$ implies exact periodicity of the field intensity along the $x$ and
$z$ directions, with periods $L_x$ and $L_z$, respectively.
Although the corresponding transverse components $k_{ymn}$ are not equally
spaced, this discretization leads to an approximate repetition of the field
pattern along the perpendicular direction $y$, which can further contribute
to crosstalk effects in configurations involving multiple parallel light sheets. In the examples considered in this work, this effect was the dominant mechanism responsible for crosstalk between adjacent light sheets.

However, this quasi-periodic behavior along the $y$ direction,
associated with the discrete plane-wave superposition,
can be substantially mitigated by appropriate choices of the design parameters.
For instance, increasing the values of $L_x$ and $L_z$ enlarges the spatial periodicities along the $x$ and $z$ directions and, as a consequence, also increases the characteristic repetition length along $y$.
An equivalent strategy consists in choosing the desired morphological function $F^{(p)}(x,z)$ such that it occupies only a fraction of the rectangular domain defined by $L_x$ and $L_z$.
In the ideal limit, the quasi-periodicity could be completely eliminated by replacing the discrete superposition of plane waves with a continuous one, involving integrals over $k_x$ and $k_z$.
The discrete formulation adopted here, however, was chosen in order to preserve the analytical and exact nature of the solution, as well as its mathematical
simplicity.
Another important point to note is that, depending on the morphological function
$F^{(p)}(x,z)$, the evaluation of the coefficients $A_{mn}^{(p)}$ through the
integral in Eq.~(\ref{Amn}) may require numerical computation, which can become
computationally demanding.
This is particularly relevant when $F^{(p)}(x,z)$ represents an image.
In such cases, an efficient and practical strategy is to employ the Fast Fourier
Transform (FFT) to compute the Fourier coefficients $A_{mn}^{(p)}$,
which significantly accelerates the construction of the analytical solution
given by Eq.~(\ref{psi}). In practice, the Fourier coefficients can be efficiently obtained through a two-dimensional FFT of the sampled morphological function, followed by the selection of the admissible $(m,n)$ indices of Eq.~(2) (see Ref.~\cite{NimaThesis}).\footnote{The material reported in this article was previously presented in Ref.~\cite{NimaThesis}.}

\section{Applying the method}

In this section, we apply the plane-wave (PW) method introduced in the previous section to the construction of structured light sheets whose envelopes are defined by prescribed images on the light-sheet planes.
To represent an image on a light sheet, the morphological function $F(x,z)$ is defined so as to reproduce a prescribed two-dimensional intensity pattern on the $(x,z)$ plane.
In practice, the gray-level distribution of the image is mapped onto the
modulus of $F(x,z)$, whereas its phase may be taken as constant or suitably designed, depending on the application.

We first consider the generation of a single structured light sheet and then extend the analysis to the case of multiple parallel structured light sheets.
In all cases, the results obtained with the PW method are directly compared with those obtained using an alternative approach based on arrays of Frozen Waves (FWs) \cite{fw1,fw2,fw3,fw4,exp1,livro1}, i.e., wave fields formed by superpositions of zero-order Bessel beams. The image is discretized along the propagation direction, and each longitudinal segment is associated with a corresponding FW reproducing the respective strip of the image \cite{Leo1,Leo2}.

Here, it is important to emphasize that, for the FW-based method, the spectral parameters $Q$ and $L$ are always chosen to be equal to the parameters $Q$ and $L_z$ employed in the PW approach, in order to ensure a fair and consistent comparison between the two methods.

The quality of the reconstructed images is quantitatively assessed using three parameters: the normalized cross-correlation (Corr), which measures the similarity between the target and reconstructed intensity patterns; the peak signal-to-noise ratio (PSNR), which quantifies reconstruction fidelity; and the structural similarity index measure (SSIM), which evaluates the preservation of structural image features.

These criteria are consistently used to compare the PW and FW approaches in all cases discussed below, showing that the PW method provides a more accurate and flexible representation of image-based structured light sheets.

Throughout this section, we assume propagation in free space, with refractive index $n_{RI}=1$, and a wavelength $\lambda = 532\,\mathrm{nm}$.
The experimental implementation and validation of the proposed approach are
presented separately in the next section.

\ 

\subsection{Single structured light sheet: theoretical results}

\begin{figure}[H]
\centering
\includegraphics[width=0.75\linewidth]{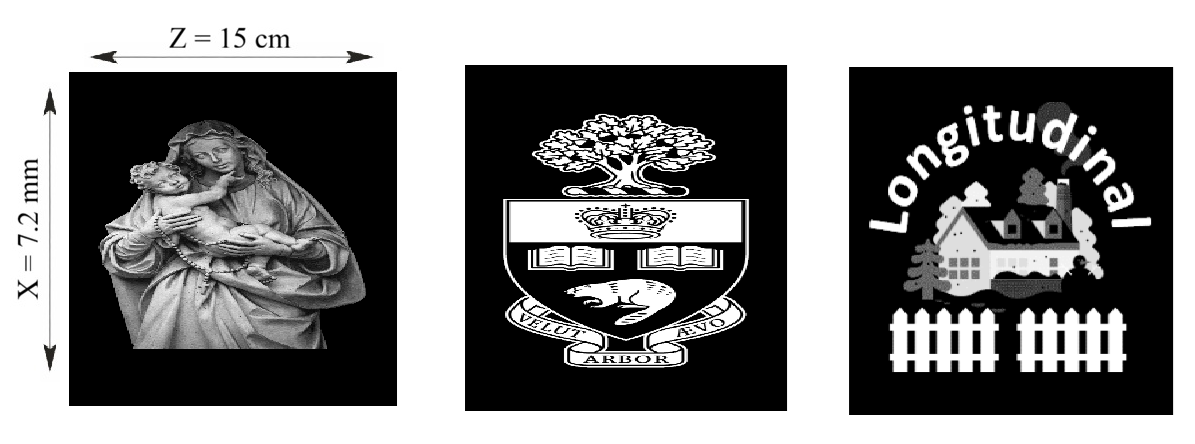}
\caption{Target images used to define the morphological function $F(x,z)$ for the single structured light-sheet demonstrations (three independent examples).}

\label{Fig1}
\end{figure}

In this subsection, we construct three structured light sheets, each one designed to reproduce one of the target images shown in Fig.~\ref{Fig1}.
Since only a single light sheet is considered in each case, the total field given by Eq.~(\ref{Psi}) reduces to a single envelope contribution, corresponding to $p=1$, i.e., $\psi^{(1)}(x,y,z)$.

The target images have a resolution of $900\times240$ pixels and occupy
spatial extensions of approximately $L_x$ and $L_z/2$ along the $x$ and $z$ directions, respectively, where $L_x$ and $L_z$ denote the characteristic transverse and longitudinal scales of the PW method and are specified later in this subsection.
Accordingly, the associated morphological functions $F(x,z)$ are defined to be nonzero only within the interval $0<z<L_z/2$, and set to zero for
$L_z/2<z<L_z$. The adopted image resolution is sufficiently high to fully exploit the spatial-frequency bandwidth available within the chosen PW spectral truncation.

The light sheets are generated using the PW method developed in Section~2, and the results are directly compared with those obtained using the methodology based on arrays of Frozen Waves (FWs).

The parameters employed in the PW method are chosen as
$L_x = 7.2\,\mathrm{mm}$, $L_z = 0.3\,\mathrm{m}$, and $Q = 0.9999\,k$. For these parameter values, the longitudinal spectral index $n$ spans $2N+1 = 113$ admissible values of $k_{z n}$. Due to the $n$-dependent transverse cutoff $M_n$, Eq.~(\ref{Psi}) results in a total of $81,514$ plane-wave terms in the superposition corresponding to a single structured light sheet.


\begin{figure}[H]
\centering
\includegraphics[width=0.75\linewidth]{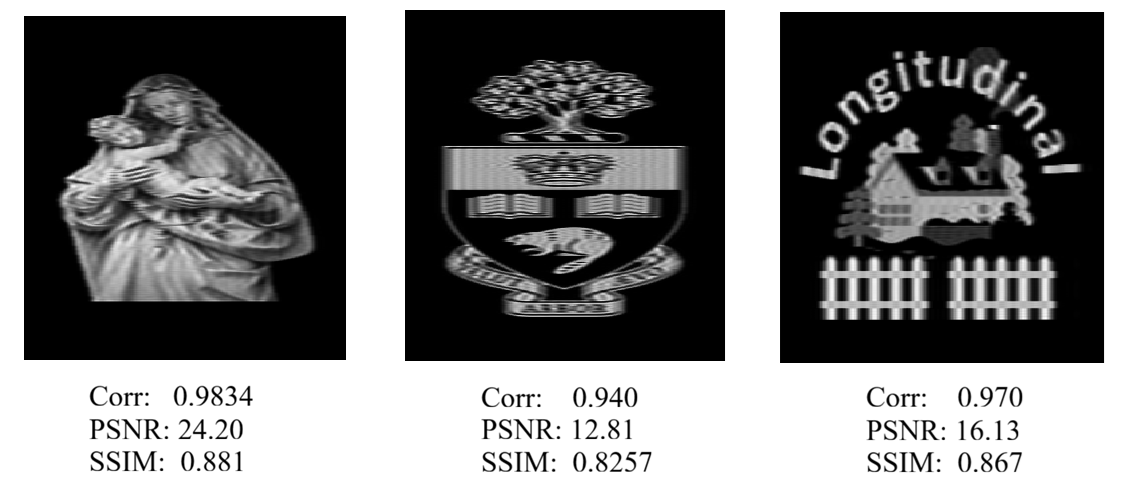}
\caption{Single structured light sheet generated with the plane-wave (PW) method. Intensity distribution evaluated on the target plane ($y=0$) for the three images of Fig.~\ref{Fig1}. The correlation, PSNR, and SSIM values are reported below each reconstruction.}

\label{Fig2}
\end{figure}

\begin{figure}[H]
\centering
\includegraphics[width=0.75\linewidth]{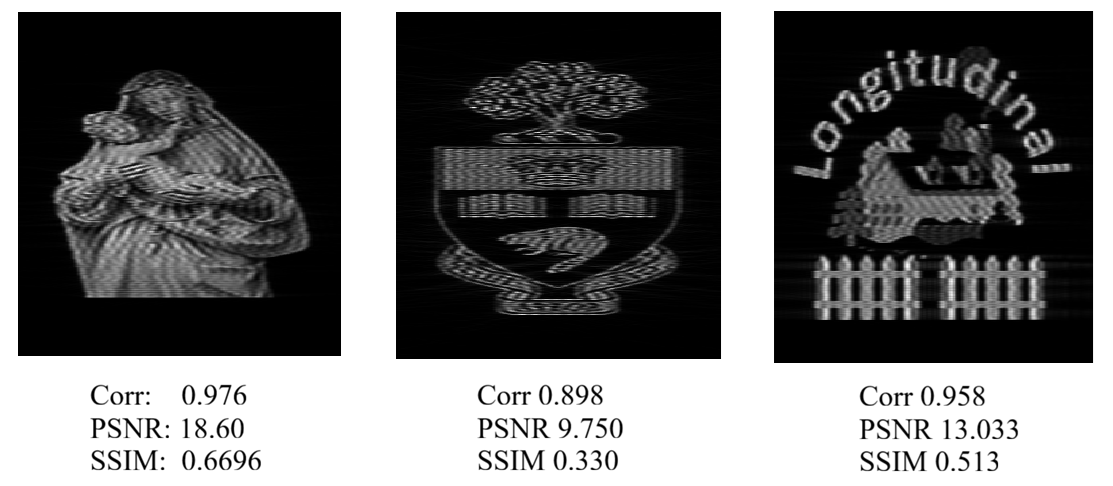}
\caption{Single structured light sheet generated with the Frozen-Wave (FW) array approach. Intensity distribution evaluated on the target plane ($y=0$) for the same images of Fig.~\ref{Fig1}. The correlation, PSNR, and SSIM values are reported below each reconstruction.}

\label{Fig3}
\end{figure}

By assigning the complex amplitudes $A_{mn}^{(1)}$ according to
Eq.~(\ref{Amn}), the analytical solution given by Eq.~(\ref{psi}) provides the structured light sheets associated with the three chosen images. Figure~\ref{Fig2} displays the intensity distributions of the resulting optical fields evaluated on the plane $y=0$ for each image obtained with the PW method.

For comparison purposes, Figure~\ref{Fig3} shows the structured light sheets obtained from the same target images using the approach based on arrays of zero-order Frozen Waves.

It is clear that the structured light sheets generated by the PW method exhibit a significantly higher fidelity to the target images.
This superiority arises from the fact that, on the plane $y=0$, the PW solution reduces to a truncated two-dimensional Fourier series, which is capable of representing arbitrary functions of two variables, such as the image-associated morphological function $F^{(1)}(x,z)$.
In contrast, arrays of zero-order FWs do not constitute a complete basis for functions of two variables, which inherently limits their ability to faithfully reproduce general image patterns on a light sheet.

\

\subsection{Multiple parallel structured light sheets: theoretical results}

\begin{figure}[H]
\centering
\includegraphics[width=0.72\linewidth]{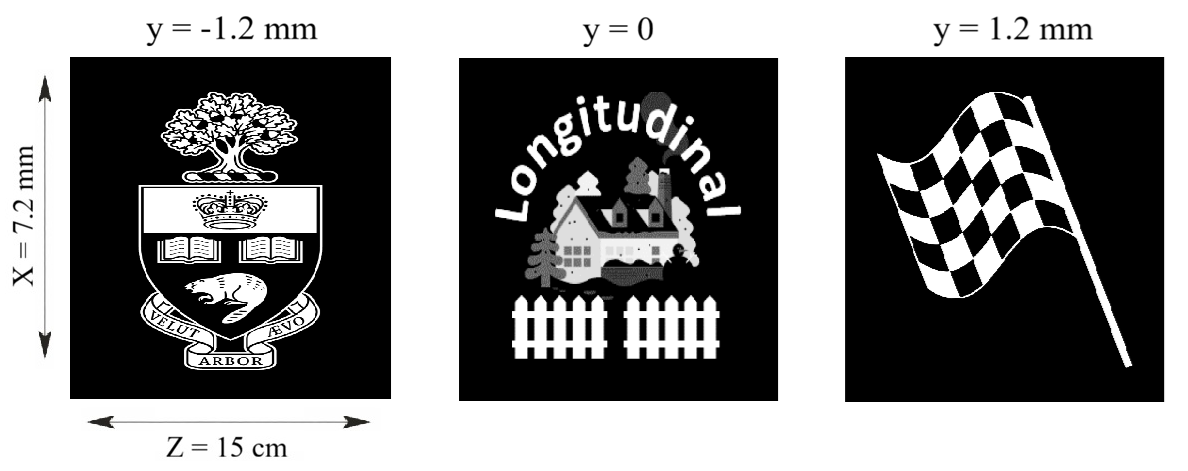}
\caption{Target images used to define the morphological functions $F^{(p)}(x,z)$ for the three parallel structured light sheets. The images are associated with the planes $y=y_p$, with $p=1,2,3$ corresponding to $y_1=-1.2\,\mathrm{mm}$, $y_2=0$, and $y_3=1.2\,\mathrm{mm}$, respectively.}
\label{Fig4}
\end{figure}

In this subsection, we consider an arrangement of three structured light sheets, parallel to each other and located on the planes $y=y_p$ $(p=1,2,3)$, with $y_1=-1.2\,\mathrm{mm}$, $y_2=0$, and $y_3=1.2\,\mathrm{mm}$.
The structured fields constructed on these light sheets have their envelopes defined by the three target images shown in Fig.~\ref{Fig4}.
Each image has a resolution of $900\times340$ pixels in the $(x,z)$ plane and occupies spatial extensions of approximately $L_x$ and $L_z/3$ along the $x$ and $z$ directions, respectively, where, again, $L_x$ and $L_z$ denote the characteristic transverse and longitudinal scales of the PW method and are specified below.

It is worth noting that, in order to mitigate crosstalk among the parallel light sheets arising from the quasi-periodic behavior of the solution given by Eq.~(\ref{Psi}) along the $y$ direction, the morphological functions
$F(x,z)$ associated with each image are defined to be nonzero only within the interval $0<z<L_z/3$, and set to zero for $L_z/3<z<L_z$.

As in the previous subsection, the array of structured light sheets generated by the PW method is directly compared with that obtained using the approach based on arrays of Frozen Waves (FWs).

For this configuration, the parameters employed in the PW method are chosen as $L_x = 7.2\,\mathrm{mm}$, $L_z = 0.45\,\mathrm{m}$, and $Q = 0.999935\,k$. With these parameter values, Eq.~(\ref{Psi}) involves $109$ allowed longitudinal spectral components $k_{z n}$, which, together with the corresponding transverse spectral components $k_{x m}$, result in a total of $63\,562$ plane-wave terms in the superposition for each of
the three structured light sheets ($p=1,2,3$).

\begin{figure}[H]
\centering
\includegraphics[width=0.72\linewidth]{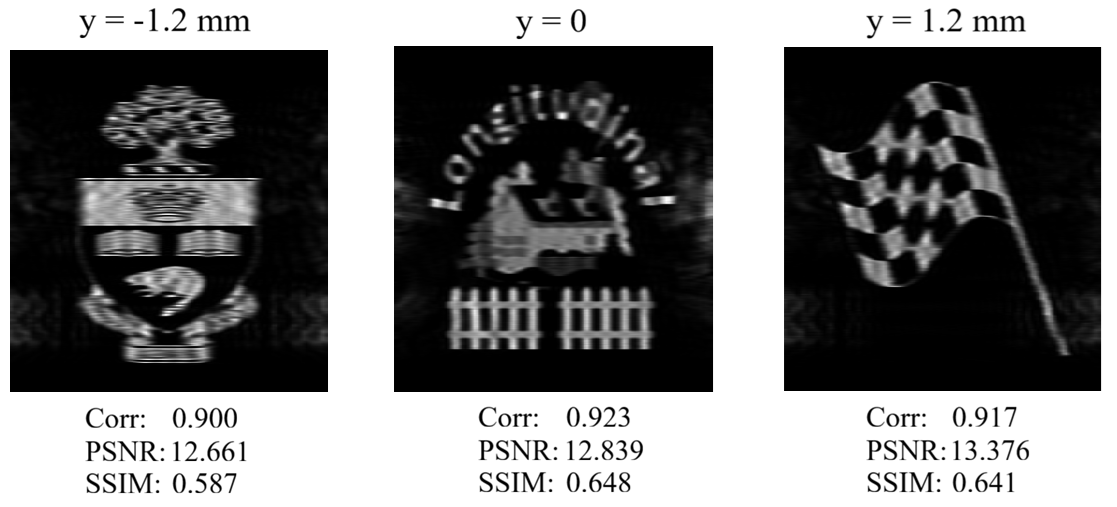}
\caption{Intensity distributions of the three parallel structured light sheets generated using the PW method, evaluated on the planes $y=y_p$ with $y_1=-1.2\,\mathrm{mm}$, $y_2=0$, and $y_3=1.2\,\mathrm{mm}$, corresponding to the target images shown in Fig.~\ref{Fig4}.}
\label{Fig5}
\end{figure}

The complex amplitudes $A_{mn}^{(p)}$ are assigned according to Eq.~(\ref{Amn}), and the analytical solution given by Eq.~(\ref{psi}) provides the structured light sheets associated with the three chosen images.
Figure~\ref{Fig5} shows the intensity distributions of the resulting light-sheet array evaluated on the planes $y=-1.2\,\mathrm{mm}$, $y=0$, and $y=1.2\,\mathrm{mm}$, as obtained with the PW method.

\begin{figure}[H]
\centering
\includegraphics[width=0.72\linewidth]{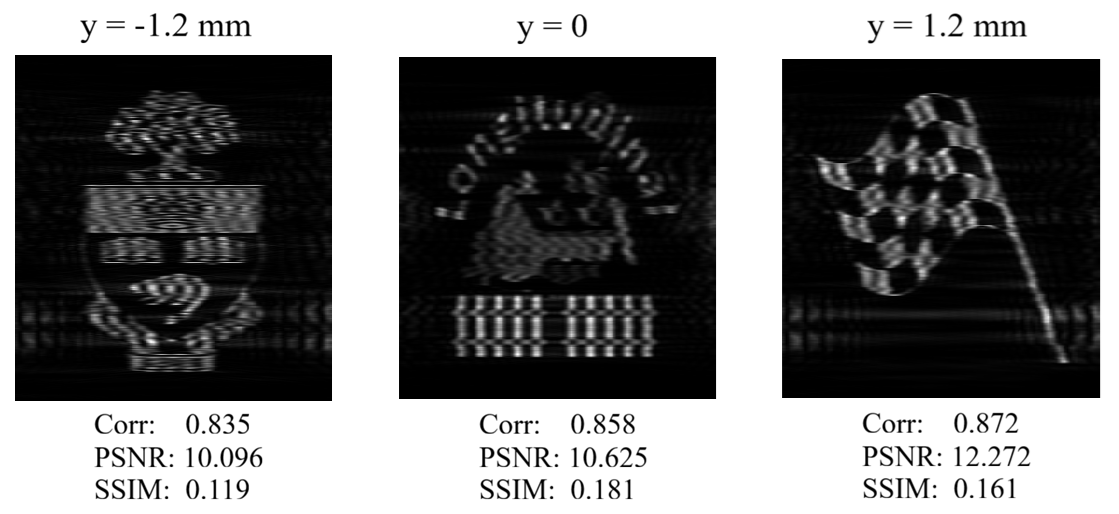}
\caption{Intensity distributions of the three parallel structured light sheets obtained using the FW-based approach, evaluated on the same planes $y=y_p$ ($y_1=-1.2\,\mathrm{mm}$, $y_2=0$, and $y_3=1.2\,\mathrm{mm}$) corresponding to the target images shown in Fig.~\ref{Fig4}.}
\label{Fig6}
\end{figure}

For comparison purposes, Fig.~\ref{Fig6} displays the corresponding array of structured light sheets generated using the approach based on arrays of zero-order Frozen Waves. Once again, the structured light sheets produced by the PW method exhibit a higher fidelity to the target images.

The results presented in this section show that the plane-wave (PW) method provides an accurate and flexible approach for the construction of image-based structured light sheets. Both single and multiple parallel light-sheet configurations are effectively described, and the direct comparison with arrays of Frozen Waves (FWs)
consistently demonstrates the superior performance of the PW approach.
The experimental implementation and validation of the proposed method are presented in the next section.

\section{Experimental validation of the method}

In this section, we present the experimental generation of the structured light sheets previously constructed through the plane-wave (PW) method.
Both the case of a single structured light sheet and the arrangement of three parallel structured light sheets are experimentally realized, using the same parameters and morphological functions adopted in the theoretical
analysis of Section~3.

\begin{figure}[ht]
\centering
\includegraphics[width=0.6\linewidth]{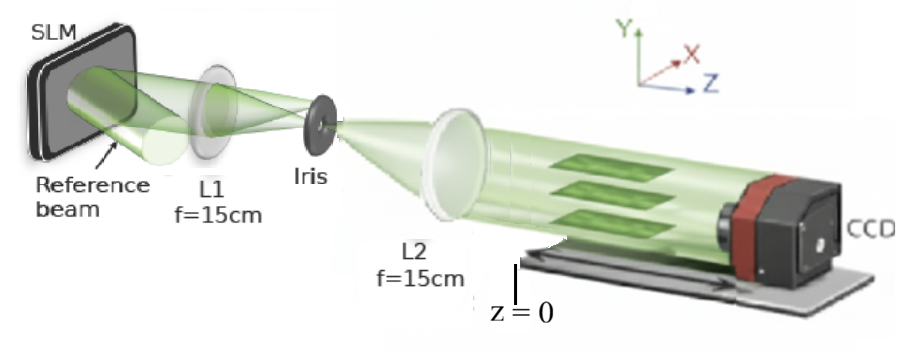}
\caption{Experimental setup for structured light-sheet generation using a phase-only SLM and a $4f$ imaging system
formed by lenses $L_1$ and $L_2$ ($f = 15\,\mathrm{cm}$).
An iris placed at the Fourier plane selects the desired diffraction order, and
the light-sheet intensity is recorded along the propagation direction $z$ using
a CCD camera mounted on a translation stage.}
\label{Fig7}
\end{figure}

The experimental implementation of the proposed approach is based on a computer-generated holography (CGH) scheme employing a reflective, phase-only spatial light modulator (SLM). A continuous-wave laser operating at a wavelength
$\lambda = 532\,\mathrm{nm}$ is spatially filtered and expanded to uniformly illuminate the SLM, onto which an off-axis hologram encoding the complex field $\Psi(x,y,z=0)$ is displayed. After reflection from the SLM, the modulated optical field is relayed through
a $4f$ optical system, where an iris placed at the Fourier plane selects the desired diffraction order, effectively suppressing unwanted spectral components.
The filtered field then propagates in free space, generating the structured light sheets predicted by the analytical model.
The intensity distributions of the generated fields are recorded at different longitudinal positions (along the $z$-direction) using a CCD camera mounted on a translation stage.
A schematic of the experimental setup is shown in Fig.~\ref{Fig7}.

Before presenting the experimental results, it is useful to briefly discuss the required size of the aperture that generates the structured light sheets.
Since the configurations considered here operate in the paraxial regime, simple ray-optics arguments can be used to estimate the minimum transverse extension of the generating aperture.
For a structured light sheet of transverse width $L_x$ that extends over a longitudinal distance $L_z$, the required aperture size along the $y$ direction at the initial plane $z=0$ can be approximately estimated as

\bb
W_y \approx 2\,\sqrt{\frac{k^2}{Q^2}-1}\,\,L_z \gg
\frac{2\pi}{\sqrt{k^2-Q^2}} \, ,
\ee
whereas the aperture size along the $x$ direction can, in general, be chosen as $W_x \gtrsim L_x$.
These conditions ensure that the angular spectrum associated with the PW solution is properly supported by the experimental system. The values of $Q$ adopted throughout this work were chosen consistently with this aperture requirement.

\begin{figure}[ht]
\centering
\includegraphics[width=0.72\linewidth]{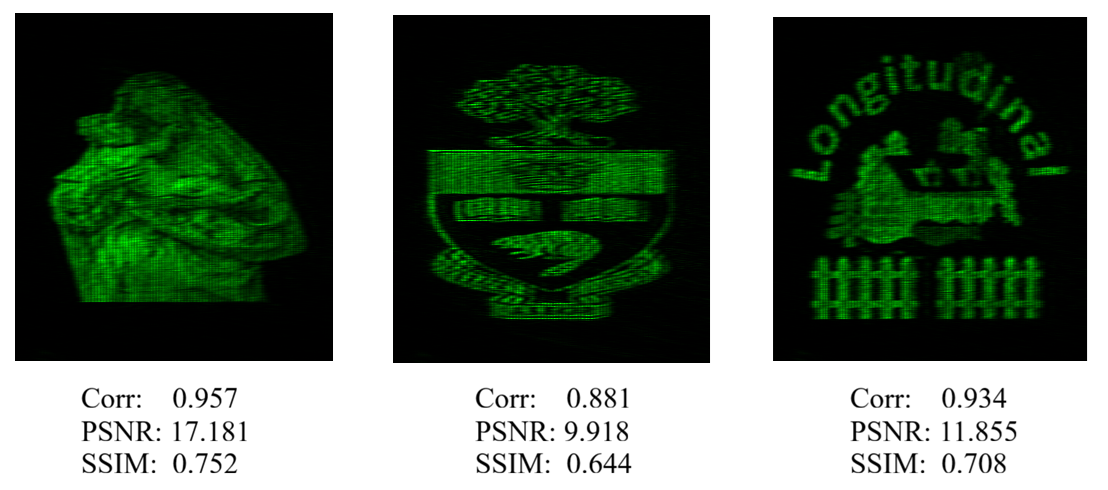}
\caption{Experimental intensity distribution of the structured light sheet generated using the PW method for the single-sheet configuration discussed in Section~3.1, recorded on the plane $y=0$, in good agreement with the theoretical prediction.}
\label{Fig8}
\end{figure}

Using the analytical solution given by Eqs.~(\ref{Psi}) and~(\ref{psi}), together with the complex amplitudes $A_{mn}^{(p)}$ obtained in Section~3, the computer-generated hologram is calculated from the field $\Psi(x,y,z=0)$ and displayed on the SLM.
Figure~\ref{Fig8} shows the experimentally generated structured light sheet corresponding to the single-sheet configuration discussed in Section~3.1.
The recorded intensity distributions are in good agreement with the theoretical predictions, confirming the ability of the PW method to faithfully imprint the desired image onto the light-sheet plane.

\begin{figure}[ht]
\centering
\includegraphics[width=0.73\linewidth]{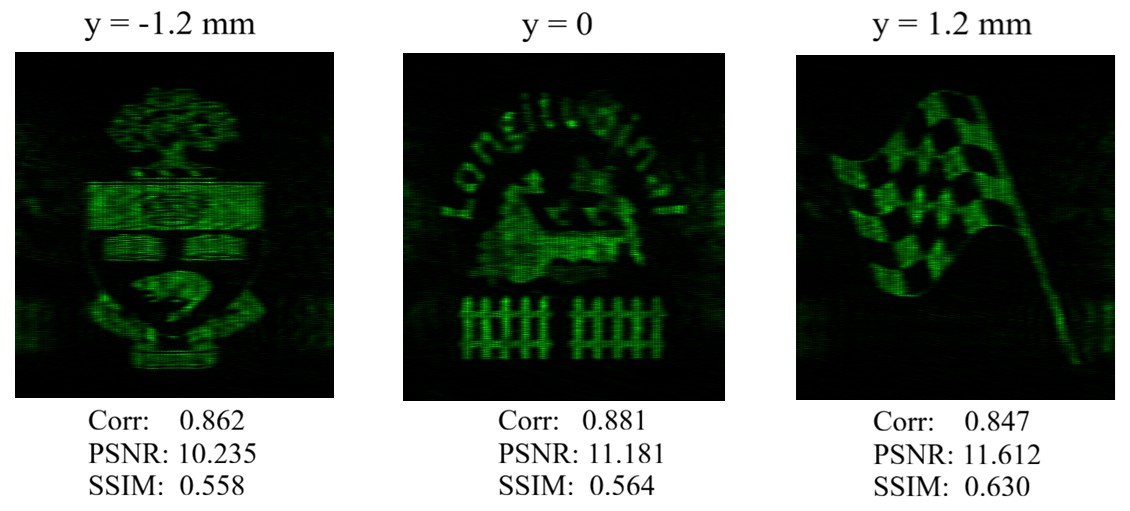}
\caption{Experimentally measured intensity distributions of the three parallel structured light sheets generated using the PW method, recorded on the planes $y=y_p$ with $y_1=-1.2\,\mathrm{mm}$, $y_2=0$, and $y_3=1.2\,\mathrm{mm}$, respectively, in good agreement with the corresponding theoretical predictions.}
\label{Fig9}
\end{figure}

Following the same procedure, the experimental generation of three parallel structured light sheets is carried out using the parameters described in Section~3.2.
Figure~\ref{Fig9} shows the experimentally measured intensity distributions on the planes $y=-1.2\,\mathrm{mm}$, $y=0$, and $y=1.2\,\mathrm{mm}$, clearly demonstrating the simultaneous formation of multiple image-based structured light sheets with controlled spatial separation.

Overall, the experimental results presented in this section provide a clear validation of the plane-wave method for the generation of structured light sheets.
Both single and multiple parallel light-sheet configurations are successfully realized in practice, in close agreement with the theoretical model, confirming the robustness, flexibility, and experimental feasibility
of the proposed approach.

\section{Conclusions}

In this work, we presented a simple, exact, and fully analytical method, based on plane-wave superposition, for the generation of structured light sheets parallel to the propagation direction, with amplitude and phase envelopes tailored on demand. By exploiting the full spectral content compatible with unidirectional propagation, the proposed approach enables the faithful synthesis of prescribed complex field distributions on prescribed light-sheet planes. The method was theoretically validated through the generation of image-based structured light sheets, considering both single-sheet and multi-sheet configurations, and consistently demonstrated higher reconstruction fidelity than an alternative strategy based on arrays of Frozen Waves (FWs).

Experimental implementations based on computer-generated holography and a phase-only spatial light modulator confirmed the practical feasibility of the approach, showing good agreement with theoretical predictions. Owing to its analytical nature, mathematical simplicity, and flexibility, the proposed framework provides a robust and versatile platform for structured light-sheet generation, with direct relevance to optical manipulation, microscopy, volumetric displays, and three-dimensional holographic imaging.


\end{document}